 \documentclass[aps,pra,reprint,10pt,twocolumn,nobalancelastpage,flushbottom,
               showpacs,showkeys,
               amsfonts,amssymb,amsmath,longbibliography,lengthcheck]{revtex4-1}
\usepackage[english]{babel}
\usepackage{dcolumn}
\usepackage{bm}
\usepackage{verbatim} % for multiple-line comment
\usepackage[mathscr]{euscript}
\usepackage{graphics,graphicx}
\usepackage[symbol]{footmisc}
\usepackage{color,xcolor}
\selectcolormodel{cmyk}
\definecolor{dark-blue}{rgb}{0,0,0.7}
\definecolor{dark-green}{rgb}{0,0.8,0}
\definecolor{dark-red}{rgb}{0.9,0,0}
\usepackage[colorlinks=true,linkcolor=blue,filecolor=dark-blue,
            citecolor=dark-green,urlcolor=blue,breaklinks=true]{hyperref}
\begin{document}

%\title{Magneto-optical effects in subwavelength nanoparticles enhanced by optically-induced magnetic resonances}

\title{Magneto-optical response enhanced by Mie resonances in nanoantennas}

\author{\firstname{Maria~G.} \surname{Barsukova$^1$}}
\author{\firstname{Alexander~S.} \surname{Shorokhov$^1$}}
\author{\firstname{Alexander~I.} \surname{Musorin$^1$}}
\author{\firstname{Dragomir~N.} \surname{Neshev$^2$}}
\author{\firstname{Yuri~S.} \surname{Kivshar$^2$}}
\author{\firstname{Andrey~A.} \surname{Fedyanin$^{1}$}} 
\email[Corresponding author:~]{fedyanin@nanolab.phys.msu.ru}
\affiliation{$^1$Faculty of Physics, Lomonosov Moscow State University, Moscow 199991, Russia\\
$^2$Nonlinear Physics Centre, Research School of Physics and Engineering, Australian National University, Canberra ACT 2601, Australia}
\date{\today}

\maketitle

%Abstract:

\textbf{Control of light by an external magnetic field is one of the important methods for modulation of its intensity and polarisation~\cite{buschow2003handbook}.  Magneto-optical effects at the nanoscale are usually observed in magnetophotonic crystals~\cite{inoue2006magnetophotonic, GotoPRL2008}, nanostructured hybrid materials~\cite{TemnovNatPhot, LukaszewNL2011, KimelNatComm2015}  or magnetoplasmonic crystals~\cite{gruninAPL2010, belotelov2011enhanced, giessenNatComm2013, maksymov2015magneto, bossini2016magnetoplasmonics}. An indirect action of an external magnetic field (e.g. through the Faraday effect) is explained by the fact that natural materials exhibit negligible magnetism at optical frequencies. However, the concept of metamaterials overcome this limitation imposed by nature by designing artificial subwavelength meta-atoms that support a strong magnetic response, usually termed as \textit{optical magnetism}, even when they are made of nonmagnetic materials~\cite{kuznetsov2016optically}.  The fundamental question is what would be the effect of the interaction between an external magnetic field and an optically-induced magnetic response of metamaterial structures. Here we make the first step toward answering this fundamental question and demonstrate the multifold enhancement of the magneto-optical response of nanoantenna lattices due to the optical magnetism.}

\begin{figure}[h]%[!ht]
\includegraphics[width=\columnwidth]{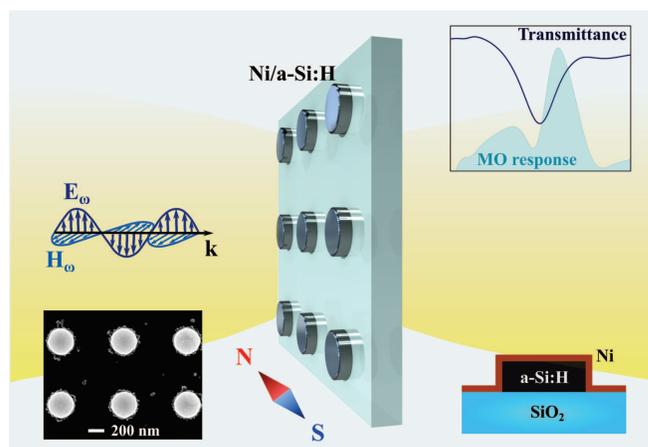}
\caption{\label{fig:1} \textbf{Schematic illustration of the enhancement of magneto-optical effects in a nanoscale structure.}  Linearly polarized light is focused on a dielectric metasurface composed of silicon nanoparticles supporting magnetic Mie-type resonances, the metasurface is covered by a thin nickel film. The sample is subjected to the action of an external magnetic field oriented perpendicular to the wave vector of the incident light. Inset shows an SEM image of the sample.}
\end{figure}

Nanophotonics is often associated with plasmonic structures made of metals such as gold or silver. However, it is known that plasmonic structures suffer from high losses of metals, heating, and incompatibility with CMOS fabrication processes. Recent developments in the nanoscale optical physics gave birth to a new branch of nanophotonics aiming at the manipulation of optically-induced Mie-type resonances in dielectric nanoparticles made of materials with high refractive indices~\cite{kuznetsov2016optically,kuznetsov2012magnetic,Evlyukhin2012,Decker:2016kr}. It has been shown recently that resonant dielectric structures offer unique opportunities for reduced dissipative losses and large resonant enhancement of both electric and magnetic fields. High-index dielectric structures can be employed as new building blocks to obtain unique functionalities such as magnetic Fano resonances~\cite{Hopkins2015, ShorokhovNL16}, highly transmittable metasurfaces~\cite{staude2013tailoring, Decker:2015iw}, and novel metadevices~\cite{kruk2016invited, Genevet:2017fk}. Here we extend the concept of high-index resonant nanophotonics to the case of \textit{magnetically active materials} and study the magneto-optical (MO) response of a dielectric metasurface covered with a thin magnetic film, as shown schematically in Fig.~\ref{fig:1}.
%----------%----------%----------%
%----------%----------%----------%
We emphasize that at the microscopic level the optical response is driven by electric dipoles, but high-index dielectric nanoparticles with this microscopic response generate effectively magnetic multipoles. Despite its geometrical simplicity, a nanoantenna is a resonant element that supports both electric and magnetic multipolar modes that can be excited efficiently optical magnetism effects. We employ such effects here and demonstrate, for the first time to our knowledge, that the optical magnetism associated with the magnetic dipole Mie resonance in a single silicon nanodisk leads to the multifold enhancement of the magneto-optical response of the hybrid magnetophotonic metasurface in comparison with a thin nickel film deposited on a flat silica substrate.

We consider metasurfaces composed of a thin nickel film deposited on top of an array of silicon nanodisks with different diameters and study optical and magneto-optical response of such hybrid planar nanostructures. The shape of the nanodisks is chosen to match the position of the lowest Mie-type resonances. Indeed, it has been shown that the electric and magnetic dipolar resonances can be overlapped spectrally by changing both height and diameter of the nanodisk~\cite{staude2013tailoring}, leading to a significant increase of the local fields inside a dielectric nanoscale structure.

For experimental studies, we use arrays of hydrogenated amorphous silicon nanodisks placed on a transparent silica substrate. The sample is covered with the 5-nm-thick nickel film in order to achieve the magneto-optical response in the external magnetic field. The SEM image of the sample is shown on Fig.~\ref{fig:1}. The disk diameter $d$ varies from 320~nm to 360~nm and the height of the nanodisks $h$ is fixed, being equal to 220~nm. The magnetic dipole Mie resonance of an isolated nanoantenna is designed to be in the visible and near-IR spectral ranges. The lattice spacing $a$ of the array is 830~nm, so that the contribution of diffraction is minimal because the phase-matching condition of the first diffraction maximum is satisfied only for the IR part of the spectrum due to the sample geometry. The contribution of higher-order resonances is negligible.

In experiment, we use normal incidence illumination and apply an external magnetic field along the sample surface. The direction of magnetization in this case is orthogonal to the direction of the light flow~(see Fig.~\ref{fig:1}). The incoming polarization is chosen to satisfy the right-hand rule: the electric field is perpendicular to both magnetization and wavevector. This configuration is known as the Voigt geometry for transmission measurements~\cite{buschow2003handbook}. The experimental geometry is similar to that used for the observation of the transverse magneto-optical Kerr effect (TMOKE), being an intensity-dependent effect observed for reflectance. An external magnetic field changes the refractive index of a magnetic medium. This action leads to a shift of the transmittance spectrum $\Delta\lambda$ relative to the non-magnetized case. The MO response of the sample can be defined as following: $\left[T(H)-T(-H)\right]/T(0),$ where $T$ is the transmittance and $H$ is the applied external magnetic field. If the spectral shift is small ($\Delta\lambda \ll \lambda$, $\lambda$ stands for the central wavelength of the resonance), the MO response is proportional to the spectral derivative of the transmittance, $\partial T/\partial\lambda$. It means that the maximum of the MO signal does not coincide with a dip or a peak in the transmission spectrum, but it corresponds to a maximal slope of the resonance curve. TMOKE is odd in the magnetization and could be observed for an oblique incidence: the sign of the effect changes with changing the angle of the incidence $\theta$ to $-\theta$, or reversing magnetic field direction. In the case of normal incidence, the MO response becomes weaker because it is  proportional to the square of the magnetization. Due to this, the MO signal in the experiment is locked-in to the doubled frequency of an alternating external magnetic field.

%----------%----------%----------%
\begin{figure}[h]
\includegraphics[width=\columnwidth]{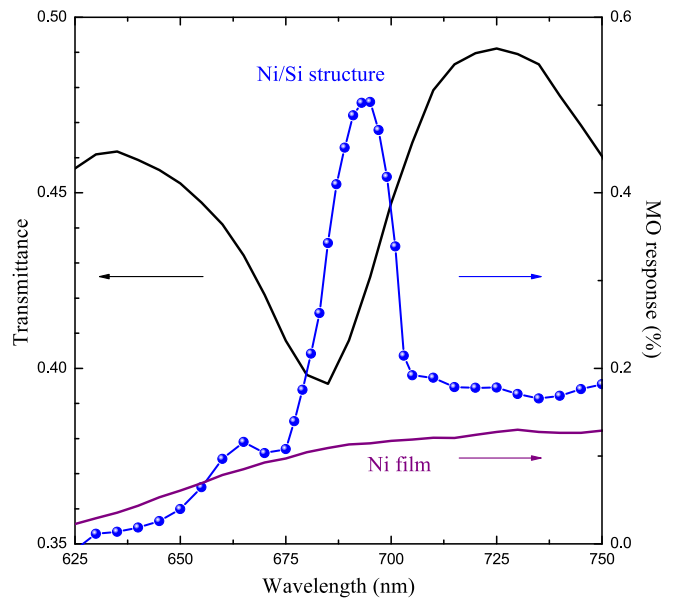}
\caption{\label{fig:2} \textbf{Experimental transmission (black, solid) and magneto-optical response (blue, circles)}. Sample comprises an array of silicon nanodisks with diameter $d = 340$~nm, height $h = 220$~nm, and lattice spacing $a = 830$~nm covered with a 5-nm-thick Ni film. The purple line illustrates the magneto-optical response of the reference nickel film.}
\end{figure}
%----------%----------%----------%

\begin{figure*}[t]
\includegraphics[width=1.8\columnwidth]{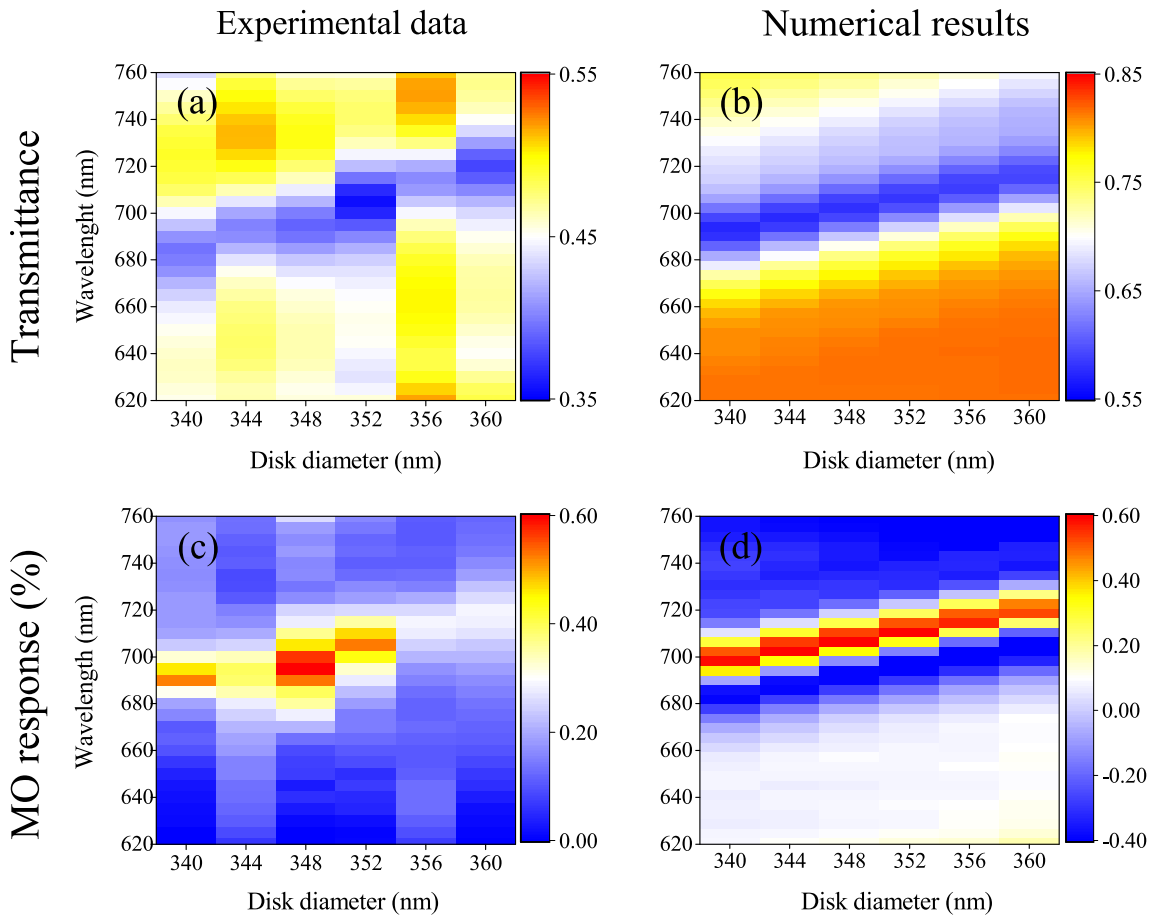}
\caption{\label{fig:3} \textbf{Control of the spectral position of the MO enhancement.} \textbf{a} and \textbf{b} illustrate experimental and numerical transmittance spectra respectively for nanodisks with different diameters. Similarly, \textbf{c} and \textbf{d} illustrate experimental and numerical magneto-optical response spectra.}
\end{figure*}

\begin{figure*}%[!ht]
\includegraphics[width=1.7\columnwidth]{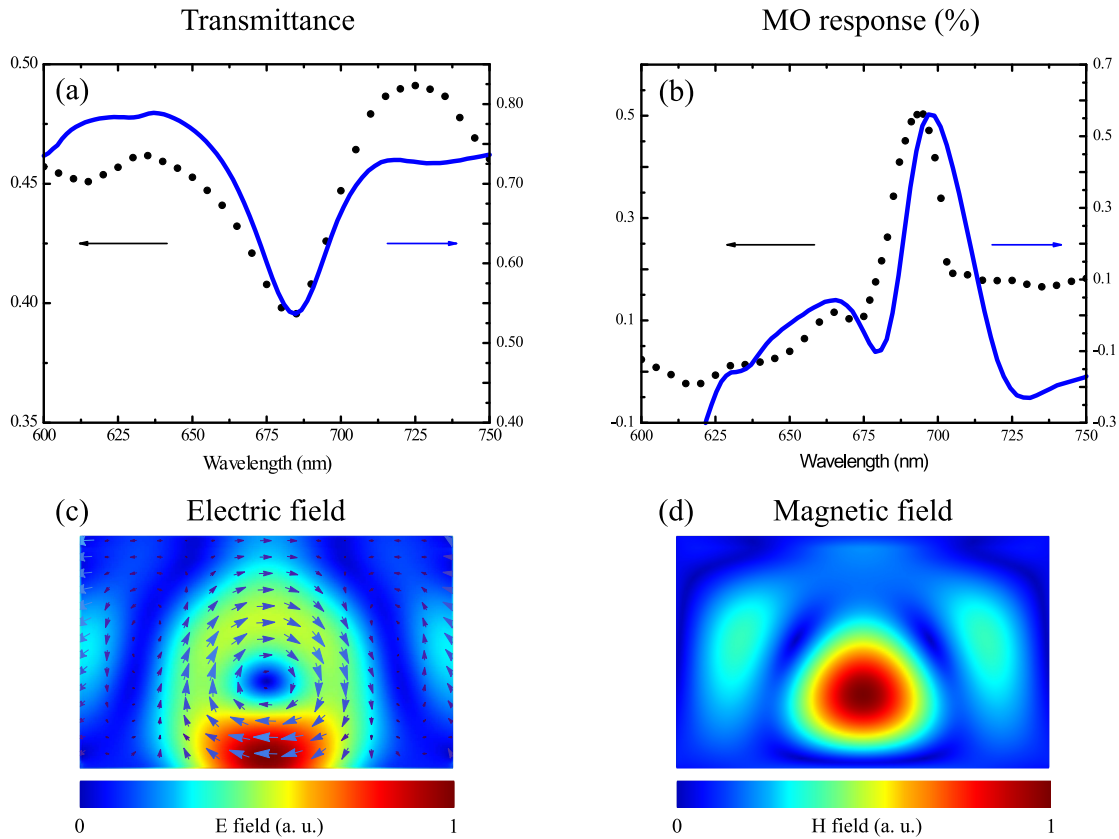}
\caption{\label{fig:4} \textbf{Comparison between experimental data and numerical results.}
Experimental (black dots) and numerical (blue solid lines) transmittance spectra \textbf{a} and magneto-optical response spectra \textbf{b} for an array of silicon nanodisks with a diameter of $d = 344$~nm, a height of $h = 220$~nm, and a lattice spacing $a = 830$~nm, covered with a 5-nm-thick nickel film. \textbf{c} and \textbf{d} illustrate the calculated electric and magnetic fields inside the nanoantenna in the vicinity of the dipole magnetic Mie resonance at $\lambda = 680$~nm.}
\end{figure*}

In order to characterize optically our MO metasurfaces, we perform measurements of the transmittance with the setup described in the Methods section (Experimental setup). Figure~\ref{fig:2} shows the experimental results for one of the samples. The sample parameters are $a=$~830~nm, $d=$~340~nm, and $h=$~220~nm. The dip in the transmission at the wavelength of 685~nm corresponds to the excitation of the magnetic dipole Mie-type resonance in a single silicon nanodisk, which has been comfirmed by numerical simulations (see Figs.~\ref{fig:3} and \ref{fig:4}). The magneto-optical response shows a resonant enhancement in the vicinity of magnetic dipole resonance of the nanodisks (blue curve). The dependence of the magneto-optical effect on the magnitude of the applied external magnetic field is shown in Supplementary Information. The magneto-optical response from the reference thin nickel film deposited on the silica substrate is illustrated in Fig.~\ref{fig:2} by a purple solid curve. The magneto-optical spectrum measured for the nickel film does not have any resonant features compared to the sample. Thus, we can easily trace the correlation between the magneto-optical enhancement and the dipole magnetic resonance excitation in the silicon nanodisks.

In order to confirm this idea, we fabricate a set of samples with increasing diameters of the silicon nanodisks. Depending on the disk diameter, the resonant wavelength is shifted: the wider the disk the redder the resonant wavelength. Figure~\ref{fig:3} shows the spectral dependencies of the transmittance (a) and the MO response (b) for samples with different diameters of the nanodisks.

%----------%----------%----------%
%\onecolumngrid

%\twocolumngrid
%----------%----------%----------%

We observe the resonance shifts from the wavelength of 680~nm (for $d=340$~nm) to 720~nm (for $d=360$~nm). The corresponding shift of the MO response is illustrated in Fig.~\ref{fig:3}\textbf{c}. To confirm our experimental data, we calculate the transmittance and MO response spectra using the finite-difference time-domain (FDTD) technique in the Lumerical FDTD Solutions software (see Methods, Numerical Simulations). In our simulations, the refractive index of glass, the complex dielectric permittivity of Si disks, and the Ni film are all taken into account.

The comparisons between experimental results and numerical simulations are shown in Fig.~\ref{fig:4}. Upper graphs illustrate experimental and numerical results for \textbf{a} transmittance and \textbf{b} magneto-optical response,  for an array of silicon nanodisks with a diameter $d = 344$~nm, a height of $h = 220$~nm, and a lattice spacing $a = 830$~nm, covered with a 5-nm-thick nickel film. The experimental and numerical spectra show quite similar behaviour, but they differ in the absolute values. This difference could be caused by the local imperfections in the form and size of nanoparticles and the roughness of the deposited Ni film, which could bring about additional scattering losses and lead to a decrease in the transmission and magneto-optical response. The lower graphs in Fig.~\ref{fig:4} demonstrate the calculated \textbf{c} electric and \textbf{d} magnetic fields inside the nanodisk in the vicinity of the transmission dip, near 680~nm. These graphs confirm the magnetic nature of the excited Mie resonance.

%----------%----------%----------%
%----------%----------%----------%

In summary, we offer the first demonstration of the magneto-optical effects enhanced by optically-induced magnetic dipole Mie resonances, manifesting a strong interaction between magnetic properties and induced \textit{optical magnetism}. Our unique findings allows for novel approaches in a magnetic control of recently reported strong nonlinear effects in nanoparticles~\cite{shcherbakov2014enhanced}, as well as the realization of the time-reversal symmetry breaking at the nanoscale for photonic topological insulators~\cite{lu2016topological}. Our results could create a new basis for novel active and nonreciprocal nanophotonic structures and metadevices, which could be tuned by external magnetic fields.

%------------------------------------------------------------------------------%
\section*{Methods}\label{sec:methods}

\textbf{Fabrication of samples.} In the capacity of a base material for our nanodisks, we use a 220~nm-thick hydrogenated amorphous silicon film deposited atop a silica substrate, using plasma-enhanced chemical vapor deposition technique in the Oxford PlasmaLab System 100. The deposition is carried out under the following conditions: 25 sccm SiH$_4$ and 475 sccm He at 250$^{\circ}$C for approximately 10 min. In order to nanostructure Si films, we use the electron-beam lithography with the subsequent reactive ion etching. For that purpose we spin-coat over the sample the negative tone ma-N-2403 electron beam resist with a thickness of 300 nm and a water-soluble anti-charging conductive polymer (ESPACER 300Z) to avoid static charging. Then we expose the resist using an electron-beam lithography system Raith 150 and remove the anti-charging layer in deionized water. The exposed film is developed in ma-D-525 for 80~s and brought to a reactive ion etching machine (Versaline LL ICP etching system). The etching process has been conducted in coupled plasma of SF$_{6}$ and CHF$_{3}$. After that, we deposite a thin nickel film on top of the sample surface using magnetron sputtering method. An SEM picture of one of the samples is shown in the insert to Fig.~\ref{fig:1}.

\textbf{Experimental setup.} Optical and magneto-optical spectra for the sample are measured with linearly polarized light from a halogen lamp, passing through a monochromator with spectral resolution of 3~nm.  After a collimating system, a 150-$\mu$m aperture and a Glan-Taylor polarizer the radiation is focused at the normal incidence to the sample surface with a beam diameter of approximately 100~$\mu$m. The light transmitted through the sample is collected and directed to a Hamamatsu photomultiplier tube via an optical fiber. The signal is detected by a lock-in amplifier at an optical chopper's frequency of 131~Hz. Magneto-optical experiments were conducted without a chopper. The reference frequency of the lock-in amplifier is set as a doubled frequency of the external alternating 300-Oe magnetic field (117~Hz). The orientation of the external magnetic field is perpendicular to the wave vector of the incident light.

\textbf{Numerical simulations.} The optical response of the sample is calculated by using the finite-difference time-domain method in the Lumerical FDTD Solutions software. In our simulations, the complex dielectric permittivity of the Ni film and Si disks as a function of frequency is taken from the standard dispersion data~\cite{palik2012handbook}. The refractive index of the glass substrate is set constant 1.45 over the whole spectral range. In order to achieve the transmittance spectrum of the sample, we model an array of silicon disks covered with a 5-nm-thick Ni film and placed on a semi-infinite glass substrate, and illuminate it at the normal incidence by a plane wave, polarized perpendicularly to the magnetization. The calculated transmission spectrum is normalized to the power of the plane wave source. We use the periodic boundary conditions along both $x$ and $y$ axis and perfectly matched layers (PML) on the top and bottom of the unit cell to prevent parasitic interference. In order to obtain the magneto-optical response spectrum, the non-diagonal complex permittivity tensor of the Ni film is taken into account. Unitary transformation makes a permittivity tensor diagonalized with eigenvalues, and the Cartesian electric field components are converted into circularly polarized light according to the permittivity tensor. The MO response is defined as following: $2*[T(H)-T(0)] /T(0)$, where $T$ is the transmittance and $H$ is the external magnetic field.

%------------------------------------------------------------------------------%

\subsection*{Appendix A: Measurements of the magneto-optical response}

In order to check the best experimental conditions, we measured the magneto-optical response as a function of the applied external magnetic field. For experimental studies we used a sample comprises an array of silicon nanodisks with a disk diameter of $d = 340$~nm, a height of $h = 220$~nm, and a period $a = 830$~nm, covered with a 5-nm-thick Ni film.

For magneto-optical measurements, we used the same experimental setup as described in the main text. The wavelength was fixed and equal to 685 nm, due to the excitation of the magnetic dipole Mie resonance in a single silicon nanodisk.

\begin{figure}
	\protect\includegraphics[width=1\columnwidth]{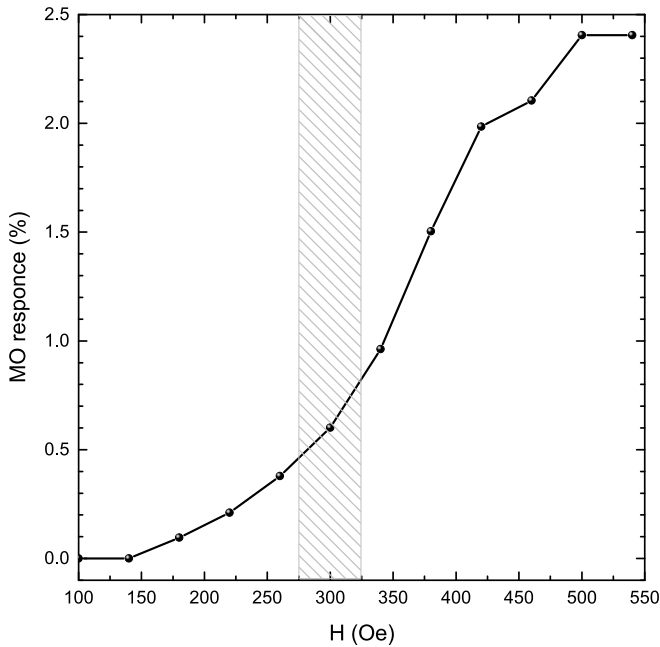}  
	\vspace{-2.5mm}
	\caption{\textbf{Dependence of the experimental magneto-optical response on the applied external magnetic field.}}
	\label{fig:figH}
\end{figure}

The magnitude of the external alternating magnetic field ($f = 117$~Hz) was varied from 100-Oe to 550-Oe. The external magnetic field was produced by a Helmholtz coil and the value of the magnetic field increased with the coil current. The magnitude of the external magnetic field was measured at the midpoint between the coils by AlphaLab Laboratory gauss-meter Model GM2. The frequency 117~Hz of the magnetic field was chosen in order to eliminate the impact of system's own mechanical resonance.  The orientation of the external magnetic field was perpendicular to the wave vector of the incident light. 

The relationship between the applied field and the magneto-optical response should not be considered as a hysteresis loop. Because we used alternating magnetic field in the experiment, at each time  $(T(+H)+T(-H))/2$ were measured.
The dependence of the magneto-optical response on the applied magnetic field trace out a curve, shown in Fig.~\ref{fig:figH}.  

\subsection*{Appendix B: Numerical simulations of the magneto-optical response}

First of all, the transmittance spectrum is calculated in the case of an absent external magnetic field. Next, the calculation of the magneto-optical phenomena in Lumerical FDTD Solutions requires three pre-steps to be done~\cite{lumerical}. 
\begin{itemize}
	\item The first one is a creation of anisotropic material. According to the experiment, we simulate a geometry when the magnetization vector is in the plane of the sample and perpendicular to the polarization of the incoming light. In our simulation, the incident plane wave propagates at normal incidence, $Z$ axis. The radiation is polarized in $Y$ direction. So, the magnetization vector is in $X$ direction. The external magnetic field makes the permittivity tensor of nickel non-diagonal:
	\begin{equation}\label{eq:EpsTMOKE}
		\hat{\varepsilon} = \left(
		\begin{array}{ccc}
			\varepsilon & 0 & 0\\
			0 & \varepsilon & ig\\
			0 & -ig & \varepsilon\\
		\end{array}\right).
	\end{equation}
	Here, $\varepsilon$ is a permittivity and $g$ is a gyration vector. The dispersrion of nickel was taken into account during the simulations for both $\varepsilon$ and $g$~\cite{krinchik1965experimental}.
	
	However, the anisotropy material in the software must be added as a diagonal matrix. Diagonalization requires eigenvalues and the unitary transformation that makes the permittivity tensor diagonal:
	\begin{equation}
		\varepsilon_{\rm{diag}} = U\varepsilon U^\dagger,
	\end{equation}
	
	where $U$ is a unitary matrix, $U^\dagger = U^{-1}$ is the complex conjugate transpose of $U$ and $\varepsilon_{\rm{diag}}$ is diagonal. $\varepsilon_{\rm{diag}}$ should be added to the material base of the software.
	
	\item The second step is a transformation of field components. This step is realized by inserting a grid attribute object, called matrix transformation. The user has to define a matrix $U$ that converts the field components from Cartesian coordinates into circularly polarized fields. In the case of magnetization in $X$ direction, the matrix $U$ is as following:
	\begin{equation}\label{eq:FrenelEquationTMOKE}
		U =
		\cfrac{1}{\sqrt{2}}
		\left(
		\begin{array}{ccc}
			\sqrt{2} & 0 & 0\\
			0 & 1 & i\\
			0 & 1 & -i\\
			\end{array}\right)
	\end{equation}
	\item The final step is a connection between a geometry object in a model and the created anisotropic material. The user has to fill in the parameter "grid attribute name" in properties of a geometric object.
\end{itemize}

Next, the transmittance spectrum is calculated in case of applying the external magnetic field. The numerical MO response, shown in the main text of the manuscript, is defined by the formula $[T(H)-T(0)] / T(0)$, where $T(H)$ and $T(0)$ is the spectral dependencies of transmittance in case of applied and absent external magnetic field, respectively.
% \newpage
% \vfill 

% \vspace{-2.5mm}

\section*{Acknowledgements}\label{sec:Acknowledgements}

The authors are grateful to M. Shcherbakov, T. Dolgova, A. Ezhov and B. Luk'yanchuk for fruitful discussions, S. Evlashin for a help with nickel film deposition, S. Dagesyan for the help with the SEM characterization, and A. Frolov for a technical assistance with the experimental setup. This work was supported by the Russian Ministry of Education and Science (\# 14.W03.31.0008) (experiment), the Russian Foundation for Basic Research (modeling) and the Australian Research Council. Calculations and measurements have been performed in Lomonosov Moscow State University. This work was performed in part at the ACT node of the Australian National Fabrication Facility, a company established under the National Collaborative Research Infrastructure Strategy to provide nano and micro-fabrication facilities for Australia’s researchers.

%------------------------------------------------------------------------------%
\section*{Author Contributions}\label{sec:Author Contributions}

M.B. conducted experiments, A. S. and A.M. helped to design and adjust an experimental setup. A.S. designed and fabricated samples, M.B. performed numerical simulations. M. B., A. S., A. M. and Y. K. wrote the first draft of the manuscript. D.N., Y.K. and A.F. guided and coordinated the entire project. All authors discussed the results and contributed to the paper writing.

%------------------------------------------------------------------------------%

\vspace{0.5cm}
\section*{Competing financial interests}
The authors declare no competing financial interests.

%------------------------------------------------------------------------------%

\end{document}